\documentclass{PoS}
\usepackage{latexsym}

\title{A High Statistics Study of Flavour-Singlet Mesons with Staggered Fermions}

\ShortTitle{A High Statistics Study of Flavour-Singlet Mesons with Staggered Fermions}

\author{UKQCD Collaboration}

\author{Eric B. Gregory,\ Craig McNeile\\
  Department of Physics and Astronomy, University of Glasgow, Glasgow G12 8QQ, UK\\
E-mail: \email{e.gregory@physics.gla.ac.uk},\ \email{c.mcneile@physics.gla.ac.uk}
}
\author{Alan C. Irving,\ \speaker{Chris Richards}\\
  Department of Mathematical Sciences, University of Liverpool, Liverpool, L69 7ZL, UK\\
  E-mail: \email{aci@liv.ac.uk},\ \email{cmr@liv.ac.uk}
}
\abstract{We present some early results from a high statistics study of the scalar and pseudoscalar singlet sectors of lattice QCD using $2+1$ flavours of Asqtad improved staggered fermions. The use of the Asqtad action has allowed us to generate an unprecedented number of configurations at 2 lattice spacings which on completion we hope will give us a significantly improved view of both the scalar and pseudoscalar singlet sectors.}

\FullConference{The XXVI International Symposium on Lattice Field Theory\\
		 July 14-19 2008\\
		 Williamsburg, Virginia, USA}

\begin{document}

\section{Introduction}
Whilst the singlet sector of QCD has been well studied using lattice QCD within the quenched approximation and even with $N_f=2$ there have been few studies using $N_f=2+1$. Computational cost is the main reason for this --- the computation of correlators for singlet mesons involves the calculation of disconnected diagrams which are inherently noisy and so require a long Monte Carlo timeseries in order to be accurately measured. 

We are engaged in a study of the scalar and pseudoscalar singlet sectors of QCD with $N_f=2+1$ flavours of ``Asqtad'' improved staggered quarks \cite{Orginos:1999cr}. The MILC ensembles have given Asqtad a strong track record in terms of physical results and they are one of the cheapest of the fermion formalisms to simulate, allowing us to build up a large number of configurations.

\section{Simulation Details}
Using the UKQCD's QCDOC \cite{Chen:2000bu} we have generated these large ensembles at 2 different lattice spacings using the one-loop tadpole-improved L\"uscher-Weisz gauge action \cite{Luscher:1985zq}
%Both have been generated with 2+1 flavours of Asqtad improved staggered fermions \cite{Orginos:1999cr} using the one-loop tadpole-improved L\"uscher-Weisz gauge action \cite{Luscher:1985zq}. 
The generation of the $\beta=7.095$ ensemble is still underway and is over halfway to completion. The lattice spacing has been obtained through a determination of $r_0/a$ from the static quark potential (taking $r_0=0.467\,\mathrm{fm}$).
\begin{table}[h]
\begin{center}
\begin{tabular}{|lll|llll|lll|}
\hline
$N_f$ & $\beta$ & $L^3\times T$ & $am_l$ & $am_s$ & ${r_0}/{a}$ & $a$ [fm] & $N_\mathrm{cfg}$ & $N_\mathrm{traj}$ & Target \\
\hline
2+1 & 6.75  & $24^3\times 64$ & 0.006   & 0.03  & 3.8122(74) & 0.12250(24) & 5375 & 32250 & 30000\\
2+1 & 7.095 & $32^3\times 64$ & 0.00775 & 0.031 & 5.059(10) & 0.09230(19) & 1911 & 11466 & 20000\\
\hline
\end{tabular}
\caption{Ensembles generated.}
\label{tab:ens}
\end{center}
\end{table}
From here on the $\beta=6.75$ ensemble will be referred to as the ``coarse'' ensemble, and the $\beta=7.095$ ensemble will be referred to as the ``fine'' ensemble.

\subsection{Algorithm}
The coarse ensemble was generated using the RHMC algorithm as formulated in \cite{Kennedy:1998cu}, with a second order leapfrog integrator. This has the benefit of being an exact algorithm, compared to the inexact R algorithm \cite{Gottlieb:1987mq} used by the MILC collaboration which has $\mathcal{O}(\delta\tau^2)$ errors, and so allows for larger integrator timesteps. The fine ensemble also used the RHMC algorithm but exploited the improvements of Clark and Kennedy \cite{Clark:2006wp} which include mass preconditioning of the light fermion kernel with the strange quark, and the $n^{th}$-root multiple pseudofermion trick. Combined these allow us to use the fourth order Omelyan integrator with the gauge fields and fermion fields on different timesteps.

%Also tuning the inversion tolerances on a per pole basis achieved a reduction in the number of conjugate gradient iterations required. As a result of these improvements we were able to use the fourth-order Omelyan integrator with the fermion fields on a timestep of $\frac{\tau}{\delta\tau}=6$ and the gauge fields on a timestep of $18$.

\subsection{Measurement}
When measuring fermionic operators using staggered fermions we must take into account both the spin and taste structure of the state in question. To this end we work in the $\mathrm{spin}\otimes\mathrm{taste}$-basis notation of Kluberg-Stern. The states which we consider are the singlet scalar $\mathbf{1}\otimes\mathbf{1}$ and the singlet pseudoscalar $\gamma_5\otimes\mathbf{1}$. In principle we could also use the $\gamma_4\gamma_5\otimes\mathbf{1}$ operator for the singlet pseudoscalar but for reasons to be discussed we use the $\gamma_5\otimes\mathbf{1}$.

The connected diagrams are measured using standard point sources, and we use the stochastic volume source method to measure the disconnected diagrams. Using $N_s$ Gaussian noise sources we compute the operators for the $\Gamma_S\otimes\Gamma_T$ state thus
\begin{equation}
\mathcal{O}_{\Gamma_S\otimes\Gamma_T}(t)=\frac{1}{N_s}\sum_{i}^{N_s}\sum_{x,x_4=t}\sum_{y,y_4=t}\mathrm{Tr}\eta^{i\dag}_{y}\Delta_{\Gamma_S\otimes\Gamma_T}M^{-1}_{yx}\eta^{i}_{x}\ ,
\end{equation}
where the $\Delta_{\Gamma_S\otimes\Gamma_T}$ covariantly and symmetrically shifts the source in the hypercube and applies the relevant Kogut-Susskind staggered phase in order to obtain the correct state. The disconnected contribution to the correlator is then calculated using
\begin{equation}
D_{\Gamma_S\otimes\Gamma_T}(\Delta t)=\langle\mathcal{O}^{\dagger}_{\Gamma_S\otimes\Gamma_T}(t)\mathcal{O}_{\Gamma_S\otimes\Gamma_T}(t+\Delta t)\rangle
\end{equation}
with the appropriate VEV subtraction for the scalar singlet.

For certain choices of $\Delta_{\Gamma_S\otimes\Gamma_T}$ (those that separate source and sink by an even number of links) we are able to apply a variance-reduction trick \cite{Venkataraman:1997xi} which allows us to use eight or fewer noise sources and still obtain a stochastic error comparable to the gauge error. This is applicable to both the $\gamma_5\otimes\mathbf{1}$ and $\mathbf{1}\otimes\mathbf{1}$ operators, and is why we choose $\gamma_5\otimes\mathbf{1}$ over $\gamma_4\gamma_5\otimes\mathbf{1}$.  However since it is only applicable to a subset we still use $N_s=32$ in order to obtain a good estimate of other operators. 

\subsection{Disconnected Statistics}
As has been mentioned previously disconnected contributions carry with them the noise of the fermionic sea and so in order to reduce statistical noise a large number of configurations is required. Even more troublesome for the $\gamma_5\otimes\mathbf{1}$ is its relation to the topological charge which has a notoriously long autocorrelation time. 

\begin{figure}
\begin{center}
\includegraphics[width=0.8\textwidth,clip]{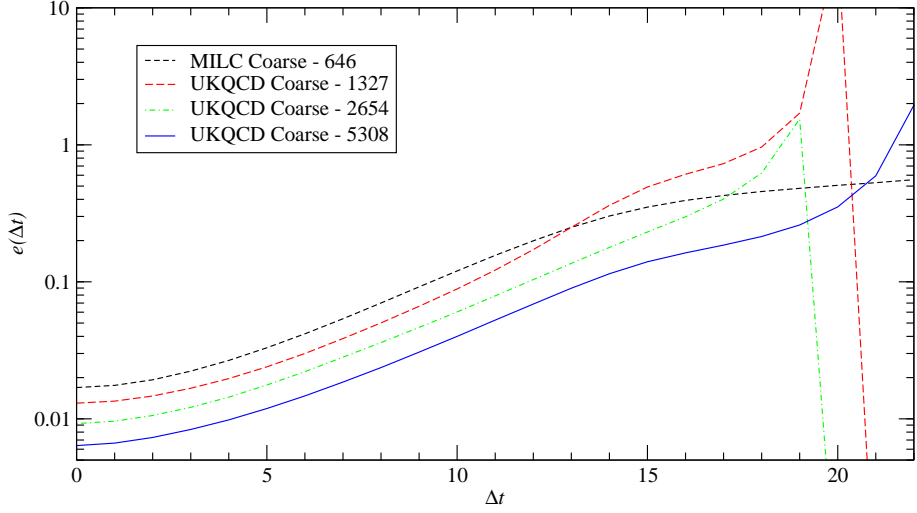}
\caption{The relative error ($e(\Delta t)\doteq\frac{\epsilon(\Delta t)}{\overline{D_{qq}}(\Delta t)}$, where $\epsilon(\Delta t)$ is the standard error on the mean of $D_{qq}(\Delta t)$) on the light-light disconnected correlators for the $\gamma_5\otimes\mathbf{1}$ (with VKVR applied, and local source and sink) for various subsets of the full UKQCD long coarse run, shown against the $\beta=6.76, m_l=0.01, m_s=0.05$ MILC coarse ensemble.}
\label{fig:errcompare}
\end{center}
\end{figure}

In Figure \ref{fig:errcompare} we show the reduction in relative error achieved from performing such a long run. This improvement should carry over directly to an improvement in the error on an estimate of the $\eta^\prime-\pi$ mass difference since the ratio of the disconnected to the connected correlator is related to this. For further details on the $\frac{D}{C}$-ratio and the statistics of the disconnected contribution the reader is referred to our methods paper \cite{Gregory:2007ev}.

\section{Scalar flavour-singlet Spectroscopy}
Whilst the heavy scalar singlet mesons ($\chi_{c0},\ \chi_{b0}$) are well understood in terms of their constituent quark content the light scalar meson sector is far from as well understood. The interpretation of the light scalar singlet states --- the $f_0$ resonances --- has eluded many experimental studies, phenomenological models and even lattice studies.

The particle data table lists five scalar singlet states with mass less than $2\ \mathrm{GeV}$ --- the $f_0(600)\ [\sigma]$, $f_0(980)$, $f_0(1370)$, $f_0(1500)$ and $f_0(1710)$. The quark model predicts only two light isoscalar $q\overline{q}$ states ($f$ and $f^\prime$) so the picture for the light isoscalar mesons is obviously more complicated. A popular conjecture is that the three heavier states are produced by a mixing of the two flavour singlet states $f$ and $f^\prime$ with the light scalar glueball. 

The scalar glueball has been well studied in quenched lattice QCD and is accepted to have a mass in the region $1.5 - 1.7\ \mathrm{GeV}$ \cite{Albanese:1987ds,Morningstar:1999rf}. However in dynamical simulations the sea quarks cause the glueball and $0^{++}$ $q\overline{q}$ interpolating operators to couple to the same states and the glueball becomes a less well defined concept. 

\subsection{Glueball Interpolating Operators}
In order to perform a mixing study it is necessary to design our operators so that they couple as strongly to the desired state as possible. In order to couple to a mainly glue state $G$ with $J^{PC}=0^{++}$ we use the following operator
\begin{equation}
\mathcal{O}_i(\mathbf{p},t)=\sum_{\vec{x}}\left(\Box^{i}_{xy}(\vec{x},t)+\Box^{i}_{yz}(\vec{x},t)+\Box^{i}_{zx}(\vec{x},t)\right)e^{\frac{2\pi i}{L}\mathbf{p}\cdot \vec{x}}
\end{equation}
where the $\Box^{i}_{kl}$ are plaquettes in the $kl$-plane with $i=0,\ldots,3$ levels of Teper blocking \cite{Teper:1987wt} applied to the gauge field, on top of 2 levels of APE smearing \cite{Albanese:1987ds} with smearing constant $c=2.5$.  We have measured these operators at $\mathbf{p}\cdot\mathbf{p}=0$ and $\mathbf{p}\cdot\mathbf{p}=1$ for the fine ensemble, and so far at $\mathbf{p}\cdot\mathbf{p}=0$ for the coarse ensemble. With our set of operators we form the $2\times 2$ correlator matrix C as
\begin{equation}
C_{ij}(\tau)=\langle \mathcal{O}^{\dag}_i(t)\mathcal{O}_j(t+\tau)\rangle - \langle\mathcal{O}_i\rangle\langle\mathcal{O}_j\rangle
\end{equation}
for the zero-momentum operators, and
\begin{equation}
C_{ij}(\tau)=\langle \mathcal{O}^{\dag}_i(t)\mathcal{O}_j(t+\tau)\rangle
\end{equation}
for the $\mathbf{p}\cdot\mathbf{p}=1$ operators, and use the variational method to obtain a set of operators with maximum projection onto the ground state. Finally we use the ratio of the variational eigenvalues $\lambda_0, \lambda_1$ to project out the remaining contamination \cite{Allton:2001sk}, obtaining our final mass estimate
\begin{equation}
\label{eq:projmass}
am_0(t)=\frac{am_{eff}(t)-\frac{\lambda_1}{\lambda_0}am_{eff}(t-1)}{1-\frac{\lambda_1}{\lambda_0}}\ \ .
\end{equation}

\subsection{Taste Symmetry Violation}
Rather than removing the doublers completely, the staggered formulation reduces the 16 doublers to 4 degenerate ``tastes'' of fermion. These are then reduced to one by taking the fourth-root of the fermion determinant. Unfortunately the taste symmetry is broken by interactions with the gauge field which leads to large mass discrepancies between hadrons of different tastes. The Asqtad action \cite{Orginos:1999cr} has improved taste-symmetry violations ($\mathcal{O}(\alpha_s^2 a^2)$), and is the most widely used staggered action. The physical validity of the fourth-rooting procedure has been called into question, but there is a large body of theoretical work confirming the validity of the resulting rooted action (see \cite{Sharpe:2006re,Kronfeld:2007ek} for reviews), although not everyone is convinced \cite{Creutz:2007yg}.

In previous analyses of the scalar non-singlet state $a_0(980)/a_0(1450)$ \cite{Aubin:2004wf, Gregory:2005yr} it was noticed that the mass obtained was significantly below the lowest allowed decay threshold $m_\pi+m_\eta$. Indeed its mass appeared to be more consistent with a $\pi\pi$ state which, due to G-parity, is forbidden as a decay channel for the scalar non-singlet. This was explained within the framework of staggered chiral perturbation theory (SChPT) as being due to unphysical multi-hadron states being introduced by the taste-split $\eta$-multiplet \cite{Prelovsek:2005rf}. The most significant of these is the $\pi\pi$ state which at large time completely dominates the $a_0$ correlator.

%In the case of the isoscalar singlet there is also the issue of unphysical multi-hadron states
%, and although the scalar singlet operators naturally couple to $\pi\pi$ states there is a $\pi\eta$ state introduced which could be mistaken for a strong $f_0(980)$ component. 
It is also the case that there are such unphysical contributions to the $f_0$ correlator, though fortunately there are explicit predictions from SChPT for these contributions to both the $f_0$ and $a_0$ correlators \cite{Bernard:2007qf} which in principle should allow us to include the contaminations in our fits.

\subsection{Results}
\begin{figure}
\begin{center}
\includegraphics[width=0.8\textwidth,clip]{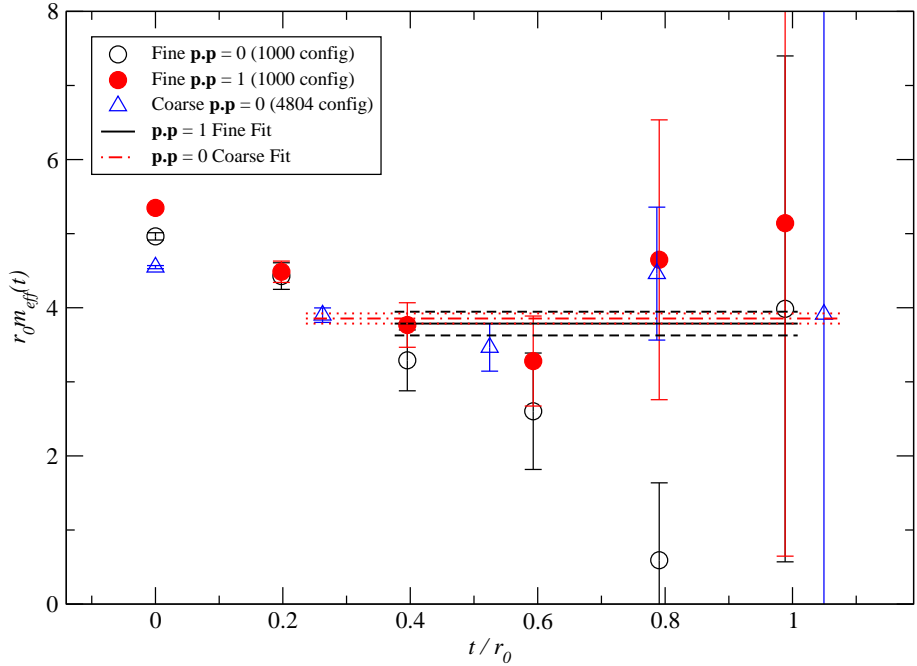}
\caption{$r_0m_\mathrm{eff}(t)$ vs $t/r_0$ for both the fine and coarse glueball operators, with both $\mathbf{p}\cdot\mathbf{p}=0$ and $1$ for the fine and $\mathbf{p}\cdot\mathbf{p}=0$ for the coarse. The effective mass has been calculated using equation \protect\ref{eq:projmass}.}
\label{fig:gballcomp}
\end{center}
\end{figure}
\begin{figure}
\begin{center}
\includegraphics[width=0.8\textwidth]{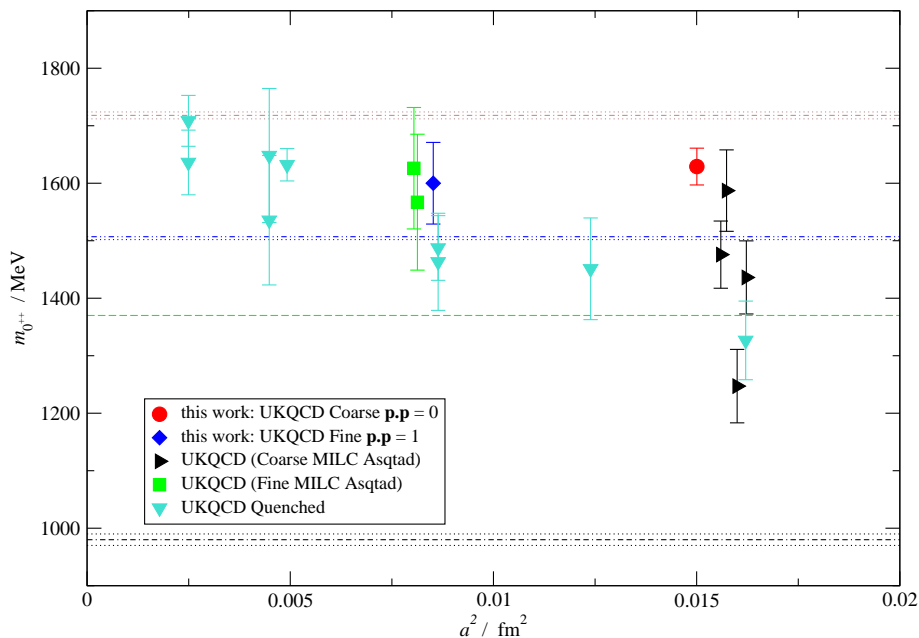}
\caption{Our measurements of the $0^{++}$ glueball mass plotted against $a^2$, with previous UKQCD analyses of the $0^{++}$ glueball on MILC ensembles and UKQCD quenched ensembles using the Wilson gauge action. The horizontal lines are at the positions of the $f_0$ resonances.}
\label{fig:gballvsa2}
\end{center}
\end{figure}

Analysis of the scalar-singlet fermionic operators is ongoing, so we present results for the glueball interpolating operators only. Our final mass estimates are obtained from the weighted average of the projected effective mass (equation \ref{eq:projmass}) over the plateau region. For the coarse ensemble this is done for the momentum-zero effective mass on 4804 configurations betweeen $t=1$ and $t=4$, and for the fine ensemble for the momentum-one effective mass on 1000 configurations between $t=2$ and $t=5$ (see Figure \ref{fig:gballcomp}). This gives masses in physical units of $M_G(0^{++})=1629(32)\ \mathrm{MeV}$ for the coarse ensemble, and $M_G(0^{++})=1600(71)\ \mathrm{MeV}$ for the fine ensemble. These are shown in the context of previous lattice determinations of the $0^{++}$ glueball mass in Figure \ref{fig:gballvsa2}.
%For the coarse ensemble this is done for the $\mathbf{p}\cdot\mathbf{p}=0$ effective mass on 4804 configurations between $t=1$ and $t=4$, and for the $\mathbf{p}\cdot\mathbf{p}=1$ values between $t=2$ and $t=5$ for 1000 configurations of the fine ensemble.  

Currently the $\mathbf{p}\cdot\mathbf{p}=0$ effective mass shows no plateau for the fine ensemble, and we have yet to measure the momentum-one operators for the coarse ensemble, but it is our intention that the weighted average be performed over both $\mathbf{p}\cdot\mathbf{p}=0$ and $1$ when this becomes possible.

\section{Pseudoscalar flavour-singlet Spectroscopy}
The light pseudoscalar flavour-singlet mesons --- $\eta$ and $\eta^\prime$ --- are particularly interesting due to the role topology plays in the mass of the latter, and the importance of disconnected diagrams in calculations involving them. It is also an important test of the validity of the fourth-root trick used with staggered fermions. It has been suggested \cite{Creutz:2007yg} that even after the continuum and chiral limits have been taken the chiral behaviour of staggered fermions will not match that of QCD. Study of the $\eta^\prime$ should help show whether or not this is the case.

The improvement in the statistical error on the disconnected operators from such high statistics has been shown in Figure \ref{fig:errcompare}. Analysis of our correlators is ongoing so we do not present any results here, but for a detailed account of our motivation and methodology the reader is referred to our exploratory study \cite{Gregory:2007ev}.

\acknowledgments
C.M.R. would like to thank Tony Kennedy and Mike Clark for useful discussions on tuning the improved RHMC algorithm.

\end{document}